\documentclass[runningheads]{llncs}
\usepackage[T1]{fontenc}
\usepackage{graphicx}
\usepackage{wasysym} % circle, half-circle symbols
\usepackage{multirow} % multiple rows in single table cell

\usepackage{todonotes}

\usepackage{tikz}
\usetikzlibrary{arrows.meta,fit,positioning,calc}

\newcommand{\tikspacing}{1em}
\newcommand{\smalltikspacing}{0.5em}
\newcommand{\tinytikspacing}{0.2em}
\tikzset{
	above of/.style={
		anchor=south, 
		above=\tikspacing of #1.north 
	},
	above of tight/.style={
		anchor=south,
		above=\smalltikspacing of #1.north
	},
	below of/.style={
		anchor=north,
		below=\tikspacing of #1.south
	},
	below of tight/.style={
		anchor=north,
		below=\smalltikspacing of #1.south
	},
	below of tighter/.style={
		anchor=north,
		below=\tinytikspacing of #1.south
	},
	right of/.style={
		anchor=west,
		right=\tikspacing of #1.east
	},
	left of/.style={
		anchor=east,
		left=\tikspacing of #1.west
	},
}

\usepackage{hyperref}
\usepackage{color}

\usepackage{cleveref}
\usepackage{xspace}
\usepackage{orcidlink}

\begin{document}
\title{MultiBallot: Verifiable and privacy-preserving E-Collecting in the Swiss setting}
\titlerunning{Verifiable and privacy-preserving E-Collecting in the Swiss setting}
\author{Florian Moser\inst{1}\orcidlink{0000-0003-2268-2367} \and
Léo Louistisserand\inst{2}\orcidlink{0009-0001-5400-8951}}

\institute{famoser GmbH, Allschwil, Switzerland \email{florian.moser@famoser.ch} 
	\and Université de Lorraine, CNRS, Inria, LORIA, Nancy, France \email{leo.louistisserand@loria.fr}}
\maketitle              % typeset the header of the contribution
\begin{abstract}
As part of the political process, citizens may participate in signature collections to influence policy changes. In Switzerland, this even results in legally binding acts, similar to an election system. In this work, we first derive a realistic setting for e-collecting in Switzerland, based on the setting established for e-voting. Then, we propose a secure protocol in this setting, achieving both privacy and verifiability under realistic trust assumptions. Notably, participation privacy is guaranteed without assuming an anonymous channel, by considering the fact that at any given point in time, many collections are active in parallel.

\end{abstract}

\section{Introduction}

In democracies, elections form the backbone of how voters decide how their government moves forward. However, besides electing their representatives, citizens often have other avenues to directly or indirectly participate in policy changes. 
This includes collection systems where citizens are able to express their political views \cite{studydatarequirementsECI2017,BR2024_ECollecting_Postulatsbericht}, and depending on the legislative basis, directly or indirectly influence policy setting. In Switzerland, as a part of the system of direct democracy, successful collections even led to a direct vote of all citizens on the topic \cite{constitution1999}. 

Overall, collecting systems have similar targets to voting systems. The system needs to make sure the voter's choices are properly recorded in the system, and their privacy is guaranteed.
For electronic voting systems, such intuitions have led to the state of the art systems that are verifiable and privacy-preserving. Verifiability guarantees the election result is auditable by third parties. Privacy ensures, despite the auditability, that the voting choice of the individual voter remains protected. 
These mechanisms are mature and employed in binding political elections \cite{moser2024e2e,heiberg2014verifiable,postspecification2026}. 

\begin{figure}[h]
	\centering
	\includegraphics[width=\textwidth]{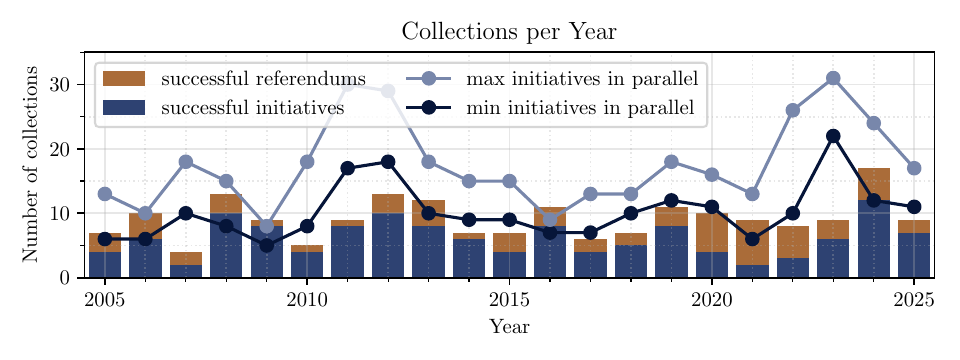}
	\caption{Swiss collections over the years  \cite{bk_initiatives}. At any point in time, at least $5$ initiatives have been in their active collection phase. \vspace{-1em}}
	\label{fig:swiss-collections}
\end{figure}

However, collection systems also have some important differences in functionality. First, a collection may start and end at any time, in particular, multiple times per year, and many collections run in parallel (see \Cref{fig:swiss-collections}). A dedicated setup process per collection, such as sending postal mail to the eligible voters \cite{postspecification2026}, is therefore unrealistic. Second, collections are long-running (many months \cite{constitution1999}). In particular, the eligibility set may change throughout the collection period (coming of age, relocation, death), while the electorate is generally assumed fixed by voting systems. Finally, signatures cannot be revoked, and always correspond to "yes" votes; there is no possibility to vote "no". Therefore, a collecting system needs to provide the stronger \textit{participation privacy}. E-voting protocols in this setting generally need to assume an anonymous channel (e.g., \cite{fujioka1992practical,arapinis2013practical,locher2016receipt}), which is considered a strong and somewhat unrealistic assumption \cite{haines2023sok}.
\vspace{-1.5em}

In academia, the security of e-collecting has not received any attention so far. Further, actual deployed e-collection systems seem to be lagging behind what is considered state-of-the-art for voting systems. 
The e-collecting system for the European Citizens Initiative (ECI) needs to comply only with a high-level directive concerning the general security of information systems in the European Union \cite{regulationECI2019,regulationSecurityEU2017}, but in particular verifiability guarantees are not a requirement. 
In Switzerland, the e-collecting system of the canton of St. Gallen, scheduled to enter production in 2026 \cite{stgallenstartecollecting2025}, seems to assume a fully trusted server\footnote{The system provider neither confirmed nor denied an attack found when the server is assumed dishonest, but instead asserted that the system is compliant with the legal basis. The cantonal legal basis omits a concrete threat model and to-be-reached properties \cite[Art. 27]{stgallenlaw2026}. Notably, in national legislation, e-voting systems must not rely on a fully trusted server \cite{veles2013}.}, and also does not provide any notion of verifiability \cite{abraxasecollectinggithub2026}.

The security of (e-)collection systems is a concern in practice. The e-collecting system for the ECI was developed as one of the recommended measures to better safeguard the personal data of supporters \cite[VI.3.1.]{studydatarequirementsECI2017}, and this same study also documents incidents of data loss and abuse happening before. In Switzerland, a report by a newspaper highlighted that many signatures (currently all given on paper) are invalid \cite{unterschriftenbeschiess2024,unterschriftenbeschiessannepolitique2024}. The suspected reason is attempted fraud by commercial signature collection firms \cite{unterschriftenbeschiess2024}, who are notably reimbursed per collected signature \cite{BuehlmannSchaub2023_ECollecting}. As a consequence, on a federal level, the Federal Chancellery has been tasked to implement e-collecting \cite{dossierforgedsignatures2025,federalecollecting2025}.

\paragraph{Contributions} 
of this works are two-fold: First, we derive a realistic setting for secure e-collecting, based on the strong setting for e-voting as used in Switzerland \cite{veles2013}. Second, in this demanding setting, we propose a protocol achieving privacy and verifiability. Notably, despite achieving participation privacy, we are able to avoid assuming an anonymous channel. The proposed protocol is therefore comparable to state-of-the-art e-voting systems in terms of security guarantees.

The protocol's core idea is the insight that many collections run in parallel (see \Cref{fig:swiss-collections}). Therefore, we are able to hide participation in some specific collection as a potential participation in any currently active collection. We are able to realize this fundamental idea using efficient cryptographic primitives with pre-existing implementations. 
\section{Related work}

We are not aware of any academic work exploring the security of e-collecting. As e-collecting systems however face similar challenges and aim for similar properties as e-voting, we explore here works written in the context of e-voting.

As argued in the introduction, participation in a collection already exposes the preference. Hence, one might aim to participate anonymously. E-voting protocols providing privacy by anonymizing the participation however need to assume an anonymous channel to cast the vote (e.g., \cite{fujioka1992practical,arapinis2013practical,locher2016receipt,lueks2020voteagain}). This is considered a strong assumption, with limited applicability in practice \cite{haines2023sok}. 

Another approach to achieve privacy is (deniable) vote updating. This approach accepts that the adversary may be able to observe some voting procedure of the voter, but allows the voter to mitigate by updating their vote later on (e.g., \cite{heiberg2014verifiable,lueks2020voteagain,muller2024devos}). However, the vote updates must still be hidden from the adversary and therefore need to rely on the anonymous channel assumption. Further, collection systems do not allow to revoke the participation, therefore once participation has been granted, revoting is no longer meaningful.

From a purely cryptographic point of view, our ballots are closest to DeVoS \cite{muller2024devos}. Concretely, DeVoS' ballots are also either a re-encryption of the previous ballot, or a fresh (proven valid) ballot from the voter. However, on the protocol level, we differ substantially. First, we do not rely on a trusted authority to hide the participation of voters. Instead, we hide participation in a specific collection by running the protocol over all collections in parallel (i.e., have multiple DeVoS-style ballots in parallel). Further, we handle mutation of eligibility of voters and collections that start and end anytime, both out of scope for DeVoS.

\section{Defining the e-collecting setting}
\label{setting}

We define an e-collecting setting, based on the extensive body of knowledge already available for e-voting. We motivate our definitions in \Cref{setting:properties} based on established notions in e-voting. The same holds for the parties and acceptable trust assumptions in \Cref{setting:parties}, which we derive out of the Swiss setting established by law \cite{veles2013}.

\subsection{Properties}
\label{setting:properties}

\newcommand{\EParticipated}{{\mathsf{Participated}}}
\newcommand{\EVoted}{{\mathsf{Voted}}}
\newcommand{\ECast}{{\mathsf{Cast}}}
\newcommand{\EVerified}{{\mathsf{Verified}}}
\newcommand{\ECounted}{{\mathsf{Counted}}}

\paragraph{End-to-End Verifiability of Participation.}

\newcommand{\LAVoters}{\mathcal{V_A}}

In internet voting, end-to-end verifiability captures the notion that the result of the election is correct. Concretely, the result of $\ECounted$ votes is composed of all votes of honest voters who $\EVerified$ successfully, a subset of the votes of honest voters that only $\EVoted$ but did not verify, and some additional votes bound by the number of adversarial voters $|\LAVoters|$. The default strong flavour of the definition further prescribes that the content of the ballot of the honest voters must be correct \cite{cortier2014election,cortier2016sok}, while the weak variant (e.g., used in \cite{cortier2025vote}) only requires this for voters who verify.

We can reuse the same approach for e-collecting. However, as participation always expresses support (i.e., the vote is always 1), we do not need to consider the adversary possibly changing votes. Therefore, we can use a weaker notion that only considers whether participation has been $\ECast$. We adapt the injective formulation of the definition \cite{cheval2023election}. Note how the definition implicitly encodes eligibility, as we understand honest voters are indeed eligible.

\begin{definition}[E2E-Verifiability of Participation]
  An e-collecting scheme provides \emph{E2E-Verifiability of Participation} if it verifies two subproperties:
  \begin{itemize}
    \item \emph{Individual verifiability}: any verified participation to a collection $C$ is part of the voting result, i.e., it holds injectively that $\EVerified(V, C) \Rightarrow \ECounted(V, C)$.
    \item \emph{Universal verifiability}: the voting result is composed solely out of at most one participation per honest voter and per attacker-controlled voter, i.e., it holds injectively that $\ECounted(V, C) \Rightarrow \ECast(V, C) \lor V \in \LAVoters$.
  \end{itemize}
\end{definition}

\paragraph{Participation privacy.}
In electronic voting, the prevalent secrecy notion is vote privacy \cite{privacy-multiset}.
It captures the intuition that the voting system should not expose more information about how individuals voted than what is known from the result.
However, participation overall in the system is not protected.
This is expressed as an indistinguishability property that states the attacker should not be able to distinguish between two settings that lead to the same election result. Concretely, in each setting, the same voters vote overall for the same multiset of votes. The adversary is however able to choose for each setting which voter votes for which voting option.

Directly reusing this definition for e-collecting is unfortunately not possible. The reason is that participation directly expresses support (i.e., the vote is always 1). Therefore, the adversary would not be able to define multiple settings, i.e., in both settings, the voters would vote for the same voting option. We can however make the definition useful again by considering multiple collection events $C_1$, $C_2$ as the different votes. 
As before, participation overall in some collection remains unprotected with this definition.

\begin{definition}[Participation Privacy]
  Given a set of voters $\mathcal{V}$, a multiset of collections $\mathcal{C}$ and two functions $f, g : \mathcal{C} \rightarrow \mathcal{V}$, we say that an e-collection scheme preserves \emph{participation privacy} if a run with $\ECast(f(C), C)$ for all $C \in \mathcal{C}$ is indistinguishable from the same run with $\ECast(g(C), C)$ for all $C \in \mathcal{C}$.
\end{definition}

\subsection{Parties}
\newcommand{\PTalliers}{{\sffamily Talliers}\xspace}
\newcommand{\PElectoralRoll}{{\sffamily Electoral Roll}\xspace}
\newcommand{\PVoter}{{\sffamily Voter}\xspace}
\newcommand{\PBulletinBoard}{{\sffamily Bulletin Board}\xspace}
\newcommand{\PAuditDevice}{{\sffamily Audit Device}\xspace}
\newcommand{\PParticipationDevice}{{\sffamily Participation Device}\xspace}
\newcommand{\PAuthChannel}{{\sffamily Authenticated Channel}\xspace}

\label{setting:parties}
At the core of the Swiss setting are the groups of control components. Each group intuitively is one distributed server, and only one of the components per group is assumed trustworthy (while it must remain unspecified which one). 
We take over this concept, and use two groups of control components: one group establishes an append-only \PBulletinBoard, and the other group shares key material and is called the \PTalliers.

During the setup of the election, an additional fully trusted setup component defines the electorate and receives key material from the control components. Then, the setup component prints out authentication and verification material and sends them over postal mail to the voters. To adapt this to the e-collecting system, we first note that e-collecting runs continuously, and not at pre-defined election events. Concretely, as a new collection may start anytime, it is tedious to each time send fresh verification material via postal mail to the voters. Further, the electorate continuously changes, rather than being fixed after a cut-off date. 
Therefore, we transfer the tasks of the Setup component to other components. First, to (continuously) define the electorate, we define the trusted \PElectoralRoll. Second, we abstract authentication of the voter from the protocol, and instead assume there is a non-repudiable \PAuthChannel from the voter to the \PBulletinBoard (e.g., using an electronic signature via an E-ID). Authenticating citizens online is an independent problem from e-collecting, and the optimal approach, including recovery or refresh of authentication material used for a prolonged time, heavily depends on existing infrastructure.

As for elections, we continue to assume the voters use their own personal devices and trust them for privacy. Concretely, we call the primary device used the \PParticipationDevice. Further, we assume an independent \PAuditDevice that the voter can use to audit their participation. While both the \PAuditDevice and the \PParticipationDevice are trusted for privacy, only one of them is assumed honest for verifiability. 

\newcommand{\chk}{$\bullet$}
\def\arraystretch{1.2}
\begin{table}
	\centering
  \begin{tabular}{lcc|r}
    		 										& \textbf{Verifiability} \hspace{1em} & \textbf{Privacy} \hspace{1em} & \hspace{1em}\textit{Difference} to e-voting setting \\
    \hline \hline
    \PBulletinBoard                & $\Circle$ & $\Circle$   &  \multirow{2}{*}{\shortstack[r]{\textit{same} as 1-out-of-n trusted \\ Control Components}}    \\
    \PTalliers                 			   & $\Circle$ &  $\Circle$ \\
    \hline
    \PElectoralRoll 				& $\CIRCLE$ &  & \multirow{2}{*}{\shortstack[r]{\textit{\color{green} weaker} than fully-trusted \\Setup and Postal Channel}}\\
    \PAuthChannel 				 & $\CIRCLE$ & & \\
    \hline
    \PParticipationDevice   & $\LEFTcircle$ & $\CIRCLE$ & \multirow{2}{*}{\shortstack[r]{\textit{\color{red} stronger} than privacy-trusted \\Voting Device}} \\
    \PAuditDevice				  & $\RIGHTcircle$ & $\CIRCLE$ & \\
    \hline \\
  \end{tabular}
  \caption{Trust assumptions for Verifiability and Privacy in the the e-collecting setting we defined, derived out of the Swiss e-voting setting \cite{veles2013}. $\CIRCLE$ for trusted, $\LEFTcircle$ / $\RIGHTcircle$ for 1-out-of-2 trusted, $\Circle$ for 1-out-of-n trusted. Stronger trust assumptions are worse.}	
  \label{setting:trust-assumptions}
\end{table}
\section{Presenting MultiBallot}

In this section, we present the MultiBallot protocol, for verifiable and privacy-preserving e-collecting. The core idea of the protocol is to hide participation in a concrete collection by making this participation indistinguishable from participations in other collections (see \Cref{proposal:idea}). MultiBallot relies only on common cryptographic primitives like public key encryption and zero-knowledge proofs, and well-understood subprotocols for distributed key generation and decryption that we recall in~\Cref{protocol:prelimaries}. Then, we present the protocol, from start to end of a collection, while also considering key rotation aspects to support indefinite running of the protocol (see \Cref{protocol:full}). Finally, we introduce an extension of MultiBallot to the hybrid setting, important for its practicality in real-world use-cases (see \Cref{protocol:hybrid}).

\subsection{Core idea}
\label{proposal:idea}

\newcommand{\definerole}[3]{\node[draw,double,minimum height=1.6em,#2] (#1) {#3}}
\newcommand{\performcalculation}[3]{\node[draw,#2] (#1) {#3}}
% args: TargetName,Source,TargetX,ArrowStyle,Content,ContentPlacement
\newcommand{\sendmessagebase}[6]{
	\coordinate[#2] (SendMessage);
	\draw[#4] (SendMessage) -- (#3|-SendMessage) node[#6]{#5};
	\coordinate (#1) at (#3|-SendMessage);
}
\newcommand{\sendpublicmessage}[5][pos=0.5,above]{\sendmessagebase{#2}{#3}{#4}{->}{#5}{#1}}

\begin{figure}[ht!]
	\centering
	\scalebox{1}{%
		\begin{tikzpicture}[
			box/.style={draw, minimum width=2em, minimum height=2em},
			frame/.style={draw, thick, inner sep=\smalltikspacing}
			]
			
			% box 1
			\node[box] (m2) {0};
			\node[box, above of tight=m2] (m1) {0};
			\node[box, below of tight=m2] (m3) {0};
			\node[frame, fit=(m1)(m2)(m3)] {};
			
			% labels
			\node[left=1em of m1] (c1) {$C_1$};
			\node[left=1em of m2] (c2) {$C_2$};
			\node[left=1em of m3] (c3) {$C_3$};
			
			% box 2
			\node[box, right=8em of m1] (r1) {0};
			\node[box, right=8em of m2] (r2) {1};
			\node[box, right=8em of m3] (r3) {0};
			\node[frame, fit=(r1)(r2)(r3)] {};
			
			% box 1 -> box 2
			\draw[->] (m1) -- node[above] {ReEnc} (r1);
			\draw[->] (m2) -- node[above] {Enc(1)} (r2);
			\draw[->] (m3) -- node[above] {ReEnc} (r3);
			
			% box 3
			\node[box, right=8em of r1] (q1) {0};
			\node[box, right=8em of r2] (q2) {1};
			\node[box, right=8em of r3] (q3) {1};
			\node[frame, fit=(q1)(q2)(q3)] {};
			
			% box 2 -> box 3
			\draw[->] (r1) -- node[above] {ReEnc} (q1);
			\draw[->] (r2) -- node[above] {ReEnc} (q2);
			\draw[->] (r3) -- node[above] {Enc(1)} (q3);
			
		\end{tikzpicture}
	}
	\caption{The ballot. For each active collection $C_1$, $C_2$ and $C_3$, there is a dedicated ciphertext. To participate, the corresponding ciphertext is replaced by an encryption of $1$. All other ciphertext are re-encrypted.}
\end{figure}
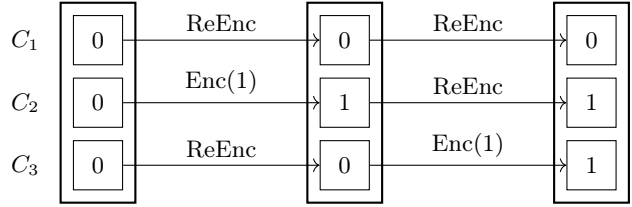

The ballot consists of a ciphertext for each active collection. To sign one or more new collections, the \PParticipationDevice encrypts 1 for these specific collections, and re-encrypts the ciphertext for all other collections. A zero-knoweldge proof ensures the \PParticipationDevice formed the ballot correctly. The new ballot then contains a fresh ciphertext for each active collection, and it is therefore indistinguishable to the adversary, which collection(s) were signed.

To enable both tallying and individual verifiability, the \PVoter choices are encrypted towards both a \PTalliers public key and a \PVoter public key. Another zero-knowledge proof ensures both encryptions behave the same (i.e., are either both a re-encryption of the previous ciphertext, or both an encryption of 1). The \PVoter can then check on an independent \PAuditDevice that the ciphertext encrypted under their public key contains the expected value. To tally, the ciphertext of all \PVoter corresponding to the same collection are aggregated, and then the \PTalliers decrypt this aggregated ciphertext.

The approach has a separate ciphertext per \PVoter and per collection, and it is thus straightforward to support eligibility mutations and arbitrary collection periods. Concretely, the \PElectoralRoll continuously adapts the whitelists of \PVoter per collection, and only \PVoter on the whitelist are allowed to replace their ballots. Further, the \PTalliers can introduce or close collections at any time, and the \PVoter can only replace ciphertext of open collections.

\subsection{Preliminaries}
\label{protocol:prelimaries}

Our protocol relies on the following common cryptographic primitives:

\newcommand{\sk}{sk}
\newcommand{\pk}{pk}
\newcommand{\KeyGen}{\mathsf{KeyGen}}
\newcommand{\Enc}{\mathsf{Enc}}
\newcommand{\Dec}{\mathsf{Dec}}
\newcommand{\CheckEnc}{\mathsf{CheckEnc}}
\newcommand{\ReRand}{\mathsf{ReRand}}
\newcommand{\Hom}{\mathsf{Hom}}

\newcommand{\ZKP}{\mathsf{ZKP}}
\newcommand{\ZKPf}[1]{\ZKP\big[#1\big]}
\newcommand{\VerifyZKP}{\mathsf{VerifyZKP}}
\newcommand{\zkpor}{||}

\begin{itemize}
	\item \emph{PKE}: A public-key encryption system with the symbols $(\sk, \pk) = \KeyGen()$, $c = \Enc(\pk, p, r)$, $p = \Dec(\sk, c)$ for ciphertext $c$, plaintext $p$ and randomness $r$. Further, we define additional non-standard functions. The function $c' = \ReRand(\pk, c, r')$ reencrypts $c$ under randomness $r'$, such that $\Dec(\sk, c) = \Dec(\sk, c')$. We write $c'' = c + c'$ to denote homomorphically add ciphertext $c$ and $c'$ such that it holds that $\Dec(\sk, c) + \Dec(\sk, c') = \Dec(\sk, c'')$. In particular, exponential El Gamal can be used to implement this functionality.
	\item \emph{ZKP}: We use zero-knowledge proofs to prove operations on ciphertext without revealing the plaintext or the encryption randomness. We denote such proofs with $\pi = \ZKPf{s}$, and it holds that if and only if statement $s$ is true, then $\VerifyZKP(\pi) = 1$. We prove statements of the form $c = \Enc(n)$ and $c' = \ReRand(c)$, where the first one asserts that $c$ is an encryption of $n$, and the second that $c'$ is a re-encryption of $c$. 
	Additionally, we compose statements with logical AND ($\land$) and OR ($\lor$). 
	Implementing this is also straightforward, in particular for DeVoS, a similar ZKP was implemented and also shown to be very efficient \cite{muller2024devos}.
\end{itemize}

\newcommand{\DKeyGen}{\mathsf{DKeyGen}}
\newcommand{\DDec}{\mathsf{DDec}}
Besides the cryptographic primitives, we assume distributed verifiable key generation and decryption as subprotocols. $\DKeyGen()$ outputs an encryption public key $\pk$. Its counterpart is $\DDec(c)$, which will, given a ciphertext $c = \Enc(\pk, m, r)$, return the corresponding plaintext $p$. These two subprotocols are executed by multiple parties, and assume at least one party is honest. Such protocols are implemented for example in Belenios \cite{cortier2019belenios}.

\subsection{Protocol}
\label{protocol:full}

\newcommand{\LVoters}{\mathcal{V}}

We divide the protocol description into multiple parts. The \textit{Setup} protocols include the \PTalliers introducing a collection, the \PVoter registering, and the \PElectoralRoll defining a whitelist of eligible voters per collection. The \textit{Participation} section describes how an eligible \PVoter participates in some of the many collections, and how the \PVoter can verify this participation. As the system is designed to run continuously, we further must account for users losing or rotating their keys, as well as mutations in eligibility (e.g., coming of age, moving to another municipality).
Finally, the \textit{Tally} describes how the participations in the collections are counted. Note that these parts of the protocol are subsequent phases per collection, but different collections may be in different phases at the same time.

We assume an established PKI in between the parties.\footnote{To authenticate the messages of the voter, for example an electronic id may be used.} All messages exchanged in the protocol are signed under this PKI, and sent to the public append-only \PBulletinBoard. The \PBulletinBoard publishes only valid messages (i.e., the signature under the PKI passes, the zero-knowledge proof of each ballot is valid, etc.). Further, we describe the \PBulletinBoard to execute all other computations that do not need any secret or random input. As it therefore operates deterministically only on public data, it is feasible to distribute the \PBulletinBoard to multiple independent parties, of which only one needs to be assumed honest. Honest parties then always act given the most recent state of the bulletin board (e.g., concerning $\pk_v$).

\paragraph{Setup}
The \PVoter registers their \PAuditDevice by generating a key pair $(\sk_v, \pk_v) = \KeyGen()$, and publishing $\pk_v$ on the bulletin board. Similarly, the \PTalliers introduce a collection by generating a distributed encryption key with $\pk_t = \DKeyGen()$, and publishing $\pk_t$ on the bulletin board. The \PElectoralRoll defines for each collection a whitelist of voters $\LVoters = \{v, \dots\}$ who are eligible.

For all registered voters $v$ in the whitelist $\LVoters$, the \PBulletinBoard generates an initial ballot for each collection. Concretely, the \PBulletinBoard encrypts $0$ into $c_t = \Enc(\pk_t, 0, r)$ for some randomness $r$, and into $c_v = \Enc(\pk_v, 0, r')$ for some randomness $r'$. Then, it creates a zero-knowledge proof attesting that $0$ has been encrypted into $c$ and $c'$, i.e., $\pi = \ZKP[c_t = \Enc(\pk_t, 0) \land  c_v = \Enc(\pk_v, 0)]$. These three values form the initial entry of the ballot $b = [(c_v, c_t, \pi)]$ corresponding to this collection, and are published on the bulletin board.

\begin{figure}[ht!]
	\centering
	\scalebox{1}{
		\begin{tikzpicture}[
			party/.style={draw, double, minimum width=2em, minimum height=2em, align=center},
			code/.style={draw, inner sep=0.5em, align=left},
			node distance=2em
			]
			
			% --- Left column (parties) ---
			\node[party] (V) {V};
			\node[party, right of=V] (AD) {AD};
			\node[party, below=of AD] (T) {T};
			\node[party, below=of T] (EL) {EL};
			\node[party, below=of EL] (BB) {BB};
			
			% --- Row 1 ---
			\node[code, right=of AD] (code1) {
				$(\sk_v, \pk_v) = \KeyGen()$
			};
			\coordinate[right of=code1, anchor=west] (endAD);
			\draw (V) -- (AD) -- (code1) -- (endAD);
			
			% --- Row 2 ---
			\node[code, right=of T] (code2) {
				$\pk_t = \DKeyGen()$
			};
			\coordinate[right of=code2] (endT);
			\draw (T) -- (code2) -- (endT);
			
			% --- Row 3 ---
			\node[code, right=of EL] (code3) {
				$\LVoters = \{v, \dots\}$
			};
			\coordinate[right of=code3] (endEL);
			\draw (EL) -- (code3) -- (endEL);
			
			% --- Vertical arrows down to BB line ---
			\draw[->] (endAD) -- ($(endAD |- BB)$);
			\draw[->] (endT) -- ($(endT |- BB)$);
			\draw[->] (endEL) -- ($(endEL |- BB)$);
			
			\coordinate (ADatBB) at ($(endAD |- BB)$);
			\node[code, right=of ADatBB, anchor=base west] (code4) {%
				$r, r' \leftarrow R$\\
				$c_t = \Enc(\pk_t, 0, r)$\\
				$c_v = \Enc(\pk_v, 0, r')$\\
				$\pi = \ZKP[c_t = c_v = \Enc(0)]$\\
				$b = [(c_t, c_v, \pi)]$
			};
			\draw (BB) -- (ADatBB) -- ++(2em,0);;
			
			% --- Frame around T, EL, BB (and their code) ---
			\coordinate[above of=T] (TSpace);
			\node[draw, dashed, inner sep=.5em,
			fit=(T)(TSpace)(endT)(EL)(endEL)(BB),
			label={[anchor=north west]north west:for every collection $C$}] {};
			
		\end{tikzpicture}
	}
	\caption{The \textit{Setup}. \PAuditDevice and \PTalliers register their corresponding public key, and the \PElectoralRoll defines the whitelist of eligible voters. Then, the initial ballot is created as an encryption of $0$. The public key of the \PAuditDevice is reused over collections.}
\end{figure}
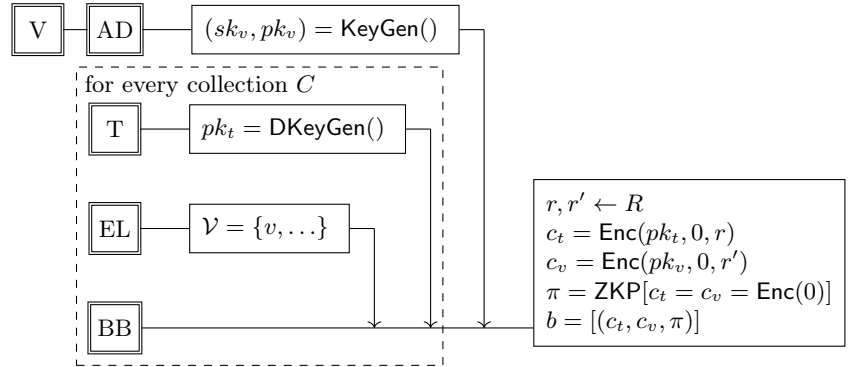

\paragraph{Participate}
To participate, the \PVoter uses their \PParticipationDevice to encrypt $1$ into $c'_t = \Enc(\pk_t, 1, r)$ for some randomness $r$, and into $c'_v = \Enc(\pk_v, 1, r')$ for some randomness $r'$. Then, the device creates an OR zero-knowledge proof attesting that \textit{either} $1$ has been encrypted into $c'_t$ and $c'_v$, \textit{or} that $c'_t$ and $c'_v$ are reencryptions of $c_v$ and $c_t$ from the last entry of the ballot $b = [\dots, (c_v, c_t, \pi)]$.	
Of course, only the first statement is true. Concretely, the proof $\pi' = \ZKP[(c'_t = \Enc(\pk_t, 1) \land  c'_v = \Enc(\pk_v, 1)) \lor (c'_t = \ReRand(\pk_t, c_t) \land  c'_v = \ReRand(\pk_v, c_v))]$ is formed. All three values are appended to the ballot, i.e., $b' =  b \| (c'_t, c'_v, \pi')$. 

To hide the participation, the \PParticipationDevice now reencrypts the other ballots. Concretely, for all other collection with the corresponding ballot $\hat{b} = (c_t, c_v, \pi)$, the ciphertext are rerandomized into $c'_t = \ReRand(\pk_t, c_t, r)$ for some randomness $r$, and into $c'_v = \ReRand(\pk_v, c_v, r')$ for some randomness $r'$. Then, the same OR zero-knowledge proof $\pi'$ is created as for participation, but now with the second statement to be true. All three values are then again appended to the respective ballot $b =  b \| (c'_t, c'_v, \pi')$. 

Finally, the \PParticipationDevice sends all $b'$ over the authenticated channel to the \PBulletinBoard. The ballots are considered valid if all ballots have received a single new entry, and if the zero-knowledge proofs are valid and therefore form a consistent chain until the initial entry.
Note that when submitting new ballots, \PVoter may also choose to sign no collection, or multiple collections at the same time. However, the zero-knowledge proofs prevent to remove the participation once it has been granted.

\begin{figure}[ht!]
	\centering
	\scalebox{1}{\begin{tikzpicture}[
			party/.style={draw, double, minimum width=2em, minimum height=2em, align=center},
			code/.style={draw, inner sep=0.5em, align=left},
			node distance=2em
			]
			
			% --- Parties row 1 ---
			\node[party] (V) {V};
			\node[party, right=of V] (VD) {PD};
			
			% --- Code row 1 ---
			\node[code, right=of VD] (code1) {%
				$r, r' \leftarrow R$\\
				$c'_t = \Enc(\pk_t, 1, r)$\\
				$c'_v = \Enc(\pk_v, 1, r')$
			};
			
			% --- Code row 2 ---
			\node[code, right=of code1] (code2) {%
				$r, r' \leftarrow R$\\
				$c'_t = \ReRand(\pk_t, c_t, r)$\\
				$c'_v = \ReRand(\pk_v, c_v, r')$
			};
			
			% --- Split point between code1 and code2 ---
			\coordinate (mid) at ($(code1.east)!0.5!(code2.west)$);
			
			% --- OR line ---
			\draw (code1.east) -- node[above] {or} (code2.west);
			\draw (V) -- (VD);
			\draw (VD) -- (code1);
			
			% --- Code 3 below middle ---
			\node[code, below=3em of mid, xshift=1.83em] (code3) {%
				$\pi' = \ZKP[c'_t = c'_v = \Enc(1) \lor (c'_t, c'_v) = \ReRand(c_t, c_v)]$\\
				$b' = b \,\|\, (c'_t, c'_v, \pi')$
			};
			
			% --- Vertical from OR line to code3 ---
			\draw (mid) -- ($(mid |- code3.north)$);
			
			% --- BB receives result from code3 ---
			\node[party, below=8em of VD] (BB) {BB};
			\draw (BB) -- ++(26em,0);
			
			\coordinate[right of=BB] (ReadInitialB);
			\draw[->, dashed] (ReadInitialB) -- node[right, yshift=-3em] {$b = \big[\_, (c_t, c_v, \pi)\big]$} ($(ReadInitialB |- VD)$);
			
			\draw[->] ($(mid |- code3.south)$) -- node[right] {$b'$} ($(mid |- BB)$);

			% --- Next row: V and AD ---
			\node[party, below=12em of V] (V2) {V};
			\node[party, right=of V2] (AD) {AD};
			
			% --- BB forwards b' down ---
			\coordinate[xshift=1em] (ReadB) at ($(mid |- BB)$);
			\draw[->, dashed] (ReadB) -- node[right] {$b'$} ($(ReadB |- AD)$);
			
			% --- Code4 block (Dec(b)) ---
			\coordinate (RecvB2) at ($(ReadB |- AD)$);
			\node[code, right=of RecvB2] (code4) {%
				$s = \Dec(\sk_v, c'_v)$ \\
				show $s$
			};
			
			% --- Flow into code4 ---
			\draw[-] (V2) -- (AD) -- (code4);
		\end{tikzpicture}
	}
	\caption{The \textit{Participation}. For every collection $C$, the \PParticipationDevice either encrypts $1$ or re-encrypts the previous ciphertext. It creates a zero-knowledge proof that the ballot is well-formed and then sends the new entry to the \PBulletinBoard. To perform the individual verification, the \PAuditDevice decrypts the ciphertext and shows the plaintext $s$ to the \PVoter.}
\end{figure}
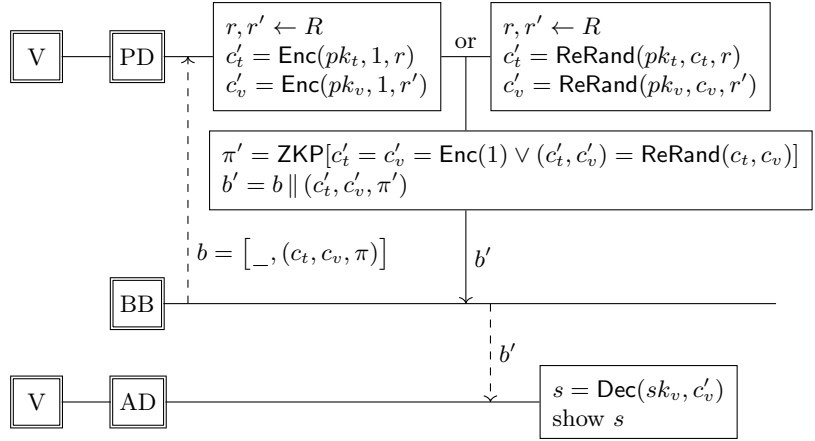

To individually verify that their ballots have been formed correctly, the \PVoter can use the registered \PAuditDevice. Concretely, the device downloads the ballots from the \PBulletinBoard, and then decrypts each last entry's $c_v$ into $\{0, 1\}$. The \PVoter then verifies whether the \PAuditDevice shows for all collections their intended choice. As any other party with access to the bulletin board, the \PAuditDevice may additionally verify the ballots, i.e., all zero-knowledge proofs are valid and form a consistent chain until the respective initial entry.

\paragraph{Rotation of keys and mutation of eligibility}
As each collection runs for an extended period of time, and as the collection system runs continuously, we need to account for key rotation. We first note that a fresh $\pk_t$ per collection is derived, and as the collection period is limited, no rotation is necessary here. 

To rotate the voter key $\pk_v$, the voter (re-)registers their \PAuditDevice. Concretely, the voter derives a new key pair $(\sk'_v, \pk'_v) = \KeyGen()$, and sends $\pk'_v$ over the authenticated channel to the bulletin board. From then on, all operations are done on $\pk'_v$ instead of $\pk_v$. This increases the complexity of the individual verification, but it remains possible, which we will detail here.

For participations created after key rotation, individual verification will work as before: The $c_v$ of the last entry of the ballot is decrypted, and results in $1$. For the other ballots, created under the old public key $\pk_v$, the decryption will fail. However, due to $\pi$, these ballots must be re-encrypted ballots, and are therefore unmodified since key-rotation. Hence, to audit these ballots, the \PAuditDevice would decrypt the last $c_v$ encrypted under the old public key $\pk_v$. 
In case the \PVoter has lost their previous secret key $\sk_v$, they will be unable to decrypt the old entries, and therefore can no longer verify participations from before $\pk'_v$ was registered. However, they can still be assured that the ballot has not been modified since the key rotation.

\begin{figure}[ht!]
	\centering
	\scalebox{1}{\begin{tikzpicture}[
			party/.style={draw, double, minimum width=2em, minimum height=2em, align=center},
			code/.style={draw, inner sep=0.5em, align=left},
			node distance=2em
			]
			
			% --- voter and BB ---
			\node[party] (V) {V};
			\node[party, right=of V] (VD) {PD};
			\node[party, below=3em of VD] (BB) {BB};
			\draw (BB) -- ++(26em,0);
			
			% --- write b and b'
			\coordinate[right of=VD] (WriteB);
			\draw[->] (WriteB) -- node[right] {$b = \big[\_, (\_, c_v, \_)\big]$} ($(WriteB |- BB)$);
			\draw[-] (V) -- (VD) -- ++(26em,0);
			
			% --- Next row: V and AD ---
			\node[party, below=8em of V] (V2) {V};
			\node[party, right=of V2] (AD) {AD};
			\node[code, right=of AD] (code1) {
				$(\sk'_v, \pk'_v) = \KeyGen()$
			};
			
			\draw[->] (code1) -- node[right] {$pk'_v$} ($(code1 |- BB)$);
			
			\coordinate[xshift=4em] (WriteB'') at ($(code1 |- VD)$);
			\draw[->] (WriteB'') -- node[right] {$b' = b \| (\_, c'_v, \_)$} ($(WriteB'' |- BB)$);
			
			\coordinate[xshift=0.5em] (ReadB'') at ($(code1.east |- BB)$);
			\draw[->, dashed] (ReadB'') -- node[right] {$b'$} ($(ReadB'' |- AD)$);
			
			\node[code, right=of code1, anchor=base west, yshift=-1.5em] (code2) {
				$s = \Dec(\sk'_v, c'_v)$\\
				if $s \neq 1$ then \\
					\hspace{0.5em} $s = \Dec(\sk_v, c_v)$\\
				show $s$
			};
			\draw[-] (V2) -- (AD) -- (code1) -- ($(code2.west |- code1)$);
			
		\end{tikzpicture}
	}
	\caption{The impact of rotating the voter encryption key on individual verification. Participations created after rotating the key can be decrypted. Otherwise, the last ciphertext from before the key rotation can be decrypted using the previous secret key.}
\end{figure}
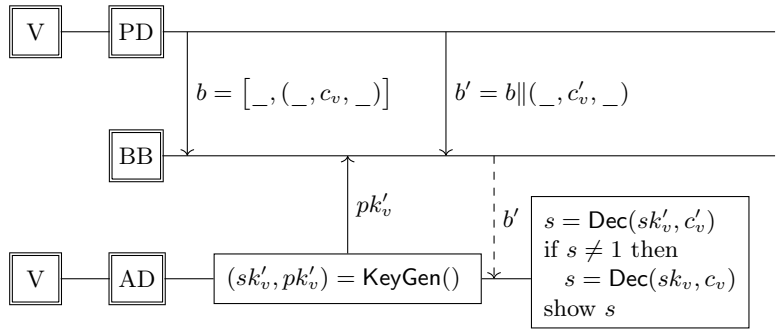

In case the \PVoter leaves an \PElectoralRoll, they will be removed from the corresponding whitelists. From then on, the \PBulletinBoard will not accept new entries for these ballots. Conversely, the \PVoter will be added to new whitelists when they enter a new \PElectoralRoll, or gain eligibility to additional collections. The \PBulletinBoard will again accept entries for the existing ballots, and create fresh initial ballots for the collections the voter has not yet been eligible for.

\paragraph{Tally}
To tally a collection, the last entries' $c_t$ are collected out of all voter's ballots. They are homomorphically added into $c_t^T$. Then, the \PTalliers execute the distributed decryption to get the sum $s = \DDec(c_t^T)$.

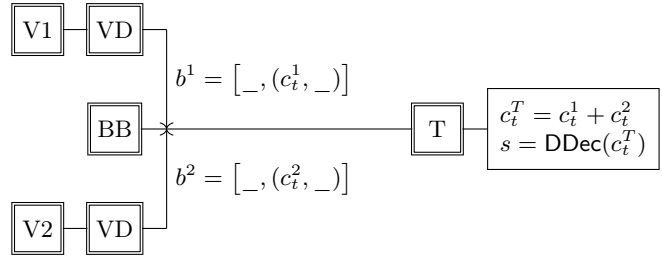
\begin{figure}[ht!]
	\centering
	\scalebox{1}{
		\begin{tikzpicture}[
			party/.style={draw, double, minimum width=2em, minimum height=2em, align=center},
			code/.style={draw, inner sep=0.5em, align=left},
			node distance=2em
			]
			
			% --- Left column (parties) ---
			\node[party] (V1) {V1};
			\node[party, right of=V1] (VD1) {VD};
			\node[party, below=of VD1] (BB) {BB};
			\node[party, below=of BB] (VD2) {VD};
			\node[party, left of=VD2] (V2) {V2};
			
			% --- VD1 arrows
			\coordinate[right of=VD1, anchor=west] (endVD1);
			\draw (V1) -- (VD1) -- (endVD1);
			\draw[->] (endVD1) -- node[right] {$b^1 = \big[\_, (c_t^1, \_)\big]$} ($(endVD1 |- BB)$);
			
			% --- VD2 arrows
			\coordinate[right of=VD2, anchor=west] (endVD2);
			\draw (V2) -- (VD2) -- (endVD2);
			\draw[->] (endVD2) -- node[right] {$b^2 = \big[\_, (c_t^2, \_)\big]$} ($(endVD2 |- BB)$);
			
			% --- tally
			\coordinate (RecvB) at ($(endVD2 |- BB)$);
			\node[party, right=of RecvB,xshift=8em] (T) {T};
			\node[code, right of=T] (code1) {%
				$c_t^T = c_t^1 + c_t^2$\\
				$s = \DDec(c_t^T)$
			};
			
			\draw (BB) -- (T) -- (code1);

		\end{tikzpicture}
	}
	\caption{The \textit{Tally}. For every collection $C$, \PTalliers aggregate the ciphertext from the respective last entry of the ballots. Then, the aggregated ciphertext is decrypted.}
\end{figure}

As the ballots are attributed to the voters, groups of ballots may be formed and decrypted separately, e.g., for statistical purposes in age groups or per electoral roll. Further, instead of just tallying at the end of the collection period, the tally may be executed regularly to learn whether a collection has already collected sufficient signatures. However, building these kinds of partial sums introduces privacy leaks, which at the worst may make voters' participation again identifiable. Techniques from differential privacy may help to reduce these kinds of side-channels, but we consider this to be out of scope for this work.

\subsection{Hybrid setting}
\label{protocol:hybrid}

As described in \Cref{setting}, hybrid systems are common in the e-collecting setting. For example, the Swiss setting will continue to support physical signatures even if an e-collecting channel becomes available. As the use of online-only systems seems very limited, the protocol is designed to work well in the hybrid setting.

\newcommand{\HC}{{\sffamily HC}\xspace}

When the \PVoter participates over the hybrid channel \HC, e.g., via signed declarations of support on paper, \HC will execute the same participation protocol as the voter would. Concretely, for the collection the voter participated in, \HC creates encryptions of $1$, and re-encrypts all the other ballots. Together with the corresponding zero-knowledge proofs, these updated ballots are sent to the \PBulletinBoard. The \PBulletinBoard accepts them as the newest ballots of the corresponding voter. The \PVoter can then execute their individual verification check to see what participations the \HC has created in their name.

If participations over \HC are auditable by third parties, these audits can continue to be executed by the \PTalliers. Concretely, for all voters for whom there is evidence that they participated over \HC, their last entry on the bulletin board can be checked to indeed encrypt $1$. Further, for all voters for whom no evidence of participation over \HC exists, the tallier may aggregate all $c_t$ of the ballot entries submitted by \HC into $c_t^\alpha$, and all $c_t$ of the respective ballot entries immediately before into $c_t^\beta$. Then, the \PTalliers decrypt $s = \DDec(c_t^\alpha-c_t^\beta)$, and they verify that $s = 0$. Note that neither audit gives the tally any additional information besides what is already known from the evidence of \HC. In particular, if the voter did not participate over \HC, the \PTalliers do not learn whether the voter participated online.

\begin{figure}[ht!]
	\centering
	\scalebox{1}{
		\begin{tikzpicture}[
			party/.style={draw, double, minimum width=2em, minimum height=2em, align=center},
			code/.style={draw, inner sep=0.5em, align=left},
			node distance=2em
			]
			
			% --- Left column (parties) ---
			\node[party] (V) {V};
			\node[party, right of=V] (VD) {VD};
			\node[party, below=of VD] (BB) {BB};
			\node[party, below=of BB] (PC) {HC};
			\node[party, below=of PC] (T) {T};
			
			% --- voter sends ballot
			\coordinate[right of=VD, anchor=west] (endVD);
			\draw (V) -- (VD) -- (endVD);
			\draw[->] (endVD) -- node[right] {$b = \big[\_, (c_t, \_)\big]$} ($(endVD |- BB)$);
			\coordinate[xshift=2em] (SendB) at ($(endVD |- BB)$);
			\draw[->, dashed] (SendB) -- node[left] {$b$} ($(SendB |- PC)$);
			
			% --- PC updates ballot
			\coordinate[xshift=1em] (endPC) at ($(SendB |- PC)$);
			\draw (PC) -- (endPC);
			\draw[->] (endPC) -- node[right] {$b' = b \| (c'_t, \_)$} ($(endPC |- BB)$);
			\draw[->] (endPC) -- node[right] {evidence $e$} ($(endPC |- T)$);
			
			% --- T checks evidence
			\coordinate[xshift=10em] (endTRecv) at ($(endPC |- T)$);
			\draw[->, dashed] ($(endTRecv |- BB)$) -- node[right,yshift=2em] {$b'$} (endTRecv);
			\node[code, right of=endTRecv, anchor=base west, yshift=-0.5em] (code1) {
				if $e$ then\\
				\hspace{0.5em}  check $1 = \DDec(c'_t)$\\
				else\\
				\hspace{0.5em}  check $0 = \DDec(c'_t - c_t)$
			};

			\draw (BB) -- ($(endPC |- BB)$) -- ($(endTRecv |- BB)$);
			\draw (T) -- ($(code1.west |- T)$);
			
		\end{tikzpicture}
	}
	\caption{Hybrid participation over another channel \HC. When the \HC creates a participation for the \PVoter, they send corresponding evidence to the \PTalliers. If the evidence indicates the \PVoter has participated over \HC, the \PTalliers verify that the last (tallied) ciphertext indeed encrypts $1$. Otherwise, the \PTalliers verify that \HC did not stuff the participation. To ease presentation, this Figure simplifies the audit procedure (we omit the aggregation of the ciphertext from multiple voters).}
\end{figure}
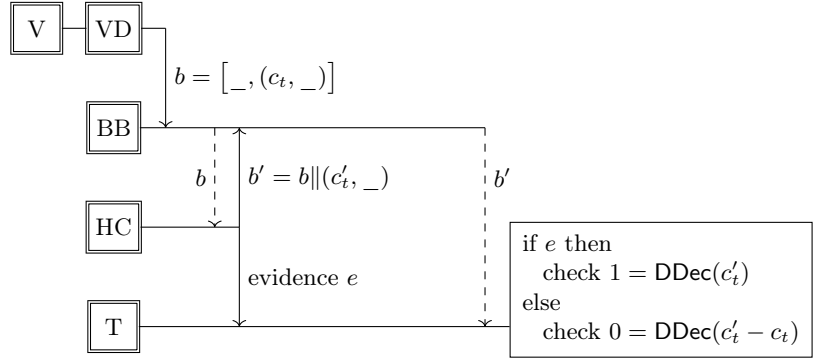

\section{Security claims}

\newcommand{\ER}{\mathsf{ER}}
\newcommand{\PD}{\mathsf{PD}}
\newcommand{\AD}{\mathsf{AD}}
\renewcommand{\HC}{\mathsf{HC}}

In this section, we give the security guarantees that our protocol offers regarding verifiability and privacy, and provide an intuition of how our claims follow. Overall, the proposal fulfils the setting as introduced in \Cref{setting}.

We consider a Dolev-Yao adversary \cite{dolev2003security} able to reorder, compose or drop messages, but that cannot break cryptography (e.g., without access to the decryption key, cannot decrypt messages). In particular, the adversary has (read-only) access to the bulletin board and may compromise some of our parties. For all security claims, we assume the \PBulletinBoard and the \PTalliers overall to be honest. Distribution of these parties (e.g., as groups of control components) is practical, hence this is not a particularly strong assumption.

\subsection{Verifiability}
\paragraph{Individual Verifiability}
When the voter $\EVerified$ their participation successfully, their participation is guaranteed to be on the bulletin board, unless the adversary compromised both the participation device $\PD$ and the audit device $\AD$. Participations on the bulletin board cannot be revoked due to the zero-knowledge proof, and therefore all will eventually be $\ECounted$.

\paragraph{Universal Verifiability}
Every participation that is $\ECounted$ is attributed to a concrete voter. For the honest voters, this means they actually $\ECast$ the participation or their $\PD$ has stuffed votes. In the hybrid setting, we need to trust the hybrid channel $\HC$ for the same purpose. 
As each participation is attributed to exactly one voter, including for attacker-controlled voters $\LAVoters$, injectivity holds. However,  overall, we need to trust the electoral roll $\ER$ to not stuff voters.

The trust assumptions however almost all disappear when audits are performed. 
A verifying voter using an honest $\AD$ would discover ballot stuffing by the $\PD$ or over the $\HC$. 
Further, if participations over $\HC$ are auditable by the \PTalliers, then they would discover any dishonest behavior by $\HC$ (both stuffing and dropping of participations). Lastly, the $\ER$ publishes the whitelists of eligible voters $\LVoters$ on the bulletin board, and $\LVoters$ can therefore be audited, too. 

\begin{figure}
	\center
	\begin{tabular}{lcc}
		& \hspace{1em} Trust assumptions \hspace{1em} & \textit{without any audits} \\
		\hline
		Individual verifiability & $\PD \vee \AD$ & ---\\
		Universal verifiability & $\PD \vee \AD$ & $\PD \wedge \HC \wedge \ER $ \\
		\hline
	\end{tabular}
	\caption{Verifiability claims for MultiBallot. We list the agents that have to be honest for each property to hold ($\PD$: participation device, $\AD$: audit device, $\HC$: hybrid channel, $\ER$: electoral roll). For universal verifiability, an honest audit device $\AD$ would discover ballot stuffing by $\HC$ or $\PD$. Further, the \PTalliers would discover a misbehaving $\HC$, and the $\ER$'s actions are public, and can therefore also be audited.}
	\label{fig:verif-claims}
\end{figure}

\subsection{Privacy}

\begin{figure}
	\center
	\begin{tabular}{l c }
		& \hspace{1em} Trust assumptions \\
		\hline
		Participation privacy  & $\PD \wedge \AD $ \\ 
		Participation privacy $\HC$ \hspace{1em}  & $\HC \wedge \AD \wedge \textrm{all talliers} $ \\ 
		\hline
	\end{tabular}
	\caption{Privacy claims for MultiBallot. We list the agents that have to be honest for each property to hold ($\PD$: participation device, $\AD$: audit device, $\HC$: hybrid channel). We denote \textit{all talliers} here explicitly, as for the audit each individual tallier will learn the evidence $e$ of $\HC$, and therefore could break the privacy of the impacted voters.}
	\label{fig:privacy-claims}
\end{figure}

The $\PD$ (and in the hybrid setting, the $\HC$) records the participation of the voter and therefore fundamentally needs to be trusted for privacy. 
Similarly, the $\AD$ will also learn of the participation, in particular as it manages $\sk_v$. 
The participation is however otherwise hidden within the ballot, as to the adversary a re-encryption is not distinguishable from an encryption of 1. The zero-knowledge proof attached to the ballot does by construction not leak which operation was executed. 
To ensure indeed all ballots reach the tally\footnote{Otherwise, there is a trivial attack where the adversary drops all except one ballot.}, we must assume that verifiability holds\footnote{Observe that this is possible under the same trust assumptions, i.e., $\PD \wedge \AD$.}, and that all honest voters indeed verify. 
Then, the homomorphic aggregation before decryption makes it impossible to learn which voters participated in which collection and which did not.

Note that when the honest \PTalliers audit the $\HC$, then all \PTalliers also learn which voters participated over $\HC$. However, they learn nothing in addition, in particular not which voters participated over the main protocol.

\section{A note about privacy}

The presented approach protects participation in some specific collection by hiding it in a potential participation in all open collections.
This follows the privacy notion in internet voting, which implicitly assumes that the voting result does not reveal how individual voters cast their vote. Already for internet voting systems, this assumption is however not always fulfilled (e.g., when the short ballot assumption does not hold \cite{kusters2012game}).

For collections, privacy leaks may be however more pronounced. As collections run over an extended period of time, each collection may have their periods of popularity, where a participation overall is likely a participation for this specific collection. If additionally the tally is executed often, these patterns of popularity are clearly observable, and may conversely also single out participations for unpopular collections. 
Further, non-participation overall leaks that no collection has been signed at all, which seems to be a semantically more important information than abstention in an election setting.

Hence, we may need to adapt a fundamentally stronger privacy notion, that even participation overall must not leak. Observe that the current proposal implicitly supports this, as voters can cast a ballot with just re-encryptions (i.e., a non-participation). However, the usability of this is likely poor, as voters would need to be assumed to regularly take a (non-)action, at least once within each tally period. To improve usability, we could transfer this re-encryption to another authority, but this would need much stronger trust assumptions\footnote{At least a hidden channel to said authority to hide the real participation for privacy.
} (e.g., as those needed for DeVoS \cite{muller2024devos}).

\bibliographystyle{splncs04}
\bibliography{references}

\end{document}